\documentclass[%
reprint,
superscriptaddress,
showpacs,
 amsmath,amssymb,
 aps,
 prl,
]{revtex4-1}

\usepackage{graphicx}
\usepackage{dcolumn}
\usepackage{bm}
\usepackage{amsmath}
\usepackage{autobreak}
\usepackage{hyperref}
\usepackage[mathlines]{lineno}
\usepackage{natbib}
\usepackage{multirow}

\begin{document}

\preprint{APS/123-QED}

\title{Constraints on Sub-GeV Dark Matter--Electron Scattering from the CDEX-10 Experiment}

\affiliation{Key Laboratory of Particle and Radiation Imaging (Ministry of Education) and Department of Engineering Physics, Tsinghua University, Beijing 100084}
\affiliation{Institute of Physics, Academia Sinica, Taipei 11529}

\affiliation{Department of Physics, Tsinghua University, Beijing 100084}
\affiliation{NUCTECH Company, Beijing 100084}
\affiliation{YaLong River Hydropower Development Company, Chengdu 610051}
\affiliation{College of Nuclear Science and Technology, Beijing Normal University, Beijing 100875}
\affiliation{College of Physics, Sichuan University, Chengdu 610065}
\affiliation{School of Physics, Peking University, Beijing 100871}
\affiliation{Department of Nuclear Physics, China Institute of Atomic Energy, Beijing 102413}
\affiliation{Sino-French Institute of Nuclear and Technology, Sun Yat-sen University, Zhuhai 519082}
\affiliation{School of Physics, Nankai University, Tianjin 300071}
\affiliation{Department of Physics, Banaras Hindu University, Varanasi 221005}
\affiliation{Department of Physics, Beijing Normal University, Beijing 100875}

\author{Z.~Y.~Zhang}
\affiliation{Key Laboratory of Particle and Radiation Imaging (Ministry of Education) and Department of Engineering Physics, Tsinghua University, Beijing 100084}
\author{L.~T.~Yang}\altaffiliation [Corresponding author: ]{yanglt@mail.tsinghua.edu.cn}
\affiliation{Key Laboratory of Particle and Radiation Imaging (Ministry of Education) and Department of Engineering Physics, Tsinghua University, Beijing 100084}
\author{Q. Yue}\altaffiliation [Corresponding author: ]{yueq@mail.tsinghua.edu.cn}
\affiliation{Key Laboratory of Particle and Radiation Imaging (Ministry of Education) and Department of Engineering Physics, Tsinghua University, Beijing 100084}

\author{K.~J.~Kang}
\affiliation{Key Laboratory of Particle and Radiation Imaging (Ministry of Education) and Department of Engineering Physics, Tsinghua University, Beijing 100084}
\author{Y.~J.~Li}
\affiliation{Key Laboratory of Particle and Radiation Imaging (Ministry of Education) and Department of Engineering Physics, Tsinghua University, Beijing 100084}
\author{M. Agartioglu}
\altaffiliation{Participating as a member of TEXONO Collaboration}
\affiliation{Institute of Physics, Academia Sinica, Taipei 11529}
\author{H.~P.~An}
\affiliation{Key Laboratory of Particle and Radiation Imaging (Ministry of Education) and Department of Engineering Physics, Tsinghua University, Beijing 100084}
\affiliation{Department of Physics, Tsinghua University, Beijing 100084}
\author{J.~P.~Chang}
\affiliation{NUCTECH Company, Beijing 100084}
\author{Y.~H.~Chen}
\affiliation{YaLong River Hydropower Development Company, Chengdu 610051}
\author{J.~P.~Cheng}
\affiliation{Key Laboratory of Particle and Radiation Imaging (Ministry of Education) and Department of Engineering Physics, Tsinghua University, Beijing 100084}
\affiliation{College of Nuclear Science and Technology, Beijing Normal University, Beijing 100875}
\author{W.~H.~Dai}
\affiliation{Key Laboratory of Particle and Radiation Imaging (Ministry of Education) and Department of Engineering Physics, Tsinghua University, Beijing 100084}
\author{Z.~Deng}
\affiliation{Key Laboratory of Particle and Radiation Imaging (Ministry of Education) and Department of Engineering Physics, Tsinghua University, Beijing 100084}

\author{C.~H.~Fang}
\affiliation{College of Physics, Sichuan University, Chengdu 610065}

\author{X.~P.~Geng}
\affiliation{Key Laboratory of Particle and Radiation Imaging (Ministry of Education) and Department of Engineering Physics, Tsinghua University, Beijing 100084}
\author{H.~Gong}
\affiliation{Key Laboratory of Particle and Radiation Imaging (Ministry of Education) and Department of Engineering Physics, Tsinghua University, Beijing 100084}
\author{Q.~J.~Guo}
\affiliation{School of Physics, Peking University, Beijing 100871}
\author{X.~Y.~Guo}
\affiliation{YaLong River Hydropower Development Company, Chengdu 610051}
\author{L. He}
\affiliation{NUCTECH Company, Beijing 100084}
\author{S.~M.~He}
\affiliation{YaLong River Hydropower Development Company, Chengdu 610051}
\author{J.~W.~Hu}
\affiliation{Key Laboratory of Particle and Radiation Imaging (Ministry of Education) and Department of Engineering Physics, Tsinghua University, Beijing 100084}

\author{H.~X.~Huang}
\affiliation{Department of Nuclear Physics, China Institute of Atomic Energy, Beijing 102413}
\author{T.~C.~Huang}
\affiliation{Sino-French Institute of Nuclear and Technology, Sun Yat-sen University, Zhuhai 519082}

\author{H.~T.~Jia}
\affiliation{College of Physics, Sichuan University, Chengdu 610065}
\author{X.~Jiang}
\affiliation{College of Physics, Sichuan University, Chengdu 610065}

\author{H.~B.~Li}
\altaffiliation{Participating as a member of TEXONO Collaboration}
\affiliation{Institute of Physics, Academia Sinica, Taipei 11529}

\author{J.~M.~Li}
\affiliation{Key Laboratory of Particle and Radiation Imaging (Ministry of Education) and Department of Engineering Physics, Tsinghua University, Beijing 100084}
\author{J.~Li}
\affiliation{Key Laboratory of Particle and Radiation Imaging (Ministry of Education) and Department of Engineering Physics, Tsinghua University, Beijing 100084}
\author{Q.~Y.~Li}
\affiliation{College of Physics, Sichuan University, Chengdu 610065}
\author{R.~M.~J.~Li}
\affiliation{College of Physics, Sichuan University, Chengdu 610065}
\author{X.~Q.~Li}
\affiliation{School of Physics, Nankai University, Tianjin 300071}
\author{Y.~L.~Li}
\affiliation{Key Laboratory of Particle and Radiation Imaging (Ministry of Education) and Department of Engineering Physics, Tsinghua University, Beijing 100084}
\author{Y.~F.~Liang}
\affiliation{Key Laboratory of Particle and Radiation Imaging (Ministry of Education) and Department of Engineering Physics, Tsinghua University, Beijing 100084}
\author {B. Liao}
\affiliation{College of Nuclear Science and Technology, Beijing Normal University, Beijing 100875}
\author{F.~K.~Lin}
\altaffiliation{Participating as a member of TEXONO Collaboration}
\affiliation{Institute of Physics, Academia Sinica, Taipei 11529}
\author{S.~T.~Lin}
\affiliation{College of Physics, Sichuan University, Chengdu 610065}
\author{S.~K.~Liu}
\affiliation{College of Physics, Sichuan University, Chengdu 610065}
\author {Y.~D.~Liu}
\affiliation{College of Nuclear Science and Technology, Beijing Normal University, Beijing 100875}
\author{Y.~Liu}
\affiliation{College of Physics, Sichuan University, Chengdu 610065}
\author {Y.~Y.~Liu}
\affiliation{College of Nuclear Science and Technology, Beijing Normal University, Beijing 100875}
\author{Z.~Z.~Liu}
\affiliation{Key Laboratory of Particle and Radiation Imaging (Ministry of Education) and Department of Engineering Physics, Tsinghua University, Beijing 100084}
\author{H.~Ma}
\affiliation{Key Laboratory of Particle and Radiation Imaging (Ministry of Education) and Department of Engineering Physics, Tsinghua University, Beijing 100084}

\author{Y.~C.~Mao}
\affiliation{School of Physics, Peking University, Beijing 100871}
\author{Q.~Y.~Nie}
\affiliation{Key Laboratory of Particle and Radiation Imaging (Ministry of Education) and Department of Engineering Physics, Tsinghua University, Beijing 100084}
\author{J.~H.~Ning}
\affiliation{YaLong River Hydropower Development Company, Chengdu 610051}
\author{H.~Pan}
\affiliation{NUCTECH Company, Beijing 100084}
\author{N.~C.~Qi}
\affiliation{YaLong River Hydropower Development Company, Chengdu 610051}
\author{J.~Ren}
\affiliation{Department of Nuclear Physics, China Institute of Atomic Energy, Beijing 102413}
\author{X.~C.~Ruan}
\affiliation{Department of Nuclear Physics, China Institute of Atomic Energy, Beijing 102413}
\author{K.~Saraswat}
\altaffiliation{Participating as a member of TEXONO Collaboration}
\affiliation{Institute of Physics, Academia Sinica, Taipei 11529}
\author{V.~Sharma}
\altaffiliation{Participating as a member of TEXONO Collaboration}
\affiliation{Institute of Physics, Academia Sinica, Taipei 11529}
\affiliation{Department of Physics, Banaras Hindu University, Varanasi 221005}
\author{Z.~She}
\affiliation{Key Laboratory of Particle and Radiation Imaging (Ministry of Education) and Department of Engineering Physics, Tsinghua University, Beijing 100084}

\author{M.~K.~Singh}
\altaffiliation{Participating as a member of TEXONO Collaboration}
\affiliation{Institute of Physics, Academia Sinica, Taipei 11529}
\affiliation{Department of Physics, Banaras Hindu University, Varanasi 221005}

\author {T.~X.~Sun}
\affiliation{College of Nuclear Science and Technology, Beijing Normal University, Beijing 100875}

\author{C.~J.~Tang}
\affiliation{College of Physics, Sichuan University, Chengdu 610065}
\author{W.~Y.~Tang}
\affiliation{Key Laboratory of Particle and Radiation Imaging (Ministry of Education) and Department of Engineering Physics, Tsinghua University, Beijing 100084}
\author{Y.~Tian}
\affiliation{Key Laboratory of Particle and Radiation Imaging (Ministry of Education) and Department of Engineering Physics, Tsinghua University, Beijing 100084}

\author {G.~F.~Wang}
\affiliation{College of Nuclear Science and Technology, Beijing Normal University, Beijing 100875}

\author{L.~Wang}
\affiliation{Department of Physics, Beijing Normal University, Beijing 100875}
\author{Q.~Wang}
\affiliation{Key Laboratory of Particle and Radiation Imaging (Ministry of Education) and Department of Engineering Physics, Tsinghua University, Beijing 100084}
\affiliation{Department of Physics, Tsinghua University, Beijing 100084}
\author{Y.~Wang}
\affiliation{Key Laboratory of Particle and Radiation Imaging (Ministry of Education) and Department of Engineering Physics, Tsinghua University, Beijing 100084}
\affiliation{Department of Physics, Tsinghua University, Beijing 100084}
\author{Y.~X.~Wang}
\affiliation{School of Physics, Peking University, Beijing 100871}

\author{H.~T.~Wong}
\altaffiliation{Participating as a member of TEXONO Collaboration}
\affiliation{Institute of Physics, Academia Sinica, Taipei 11529}
\author{S.~Y.~Wu}
\affiliation{YaLong River Hydropower Development Company, Chengdu 610051}
\author{Y.~C.~Wu}
\affiliation{Key Laboratory of Particle and Radiation Imaging (Ministry of Education) and Department of Engineering Physics, Tsinghua University, Beijing 100084}
\author{H.~Y.~Xing}
\affiliation{College of Physics, Sichuan University, Chengdu 610065}
\author{R.~Xu}
\affiliation{Key Laboratory of Particle and Radiation Imaging (Ministry of Education) and Department of Engineering Physics, Tsinghua University, Beijing 100084}
\author{Y.~Xu}
\affiliation{School of Physics, Nankai University, Tianjin 300071}
\author{T.~Xue}
\affiliation{Key Laboratory of Particle and Radiation Imaging (Ministry of Education) and Department of Engineering Physics, Tsinghua University, Beijing 100084}

\author{Y.~L.~Yan}
\affiliation{College of Physics, Sichuan University, Chengdu 610065}
\author{C.~H.~Yeh}
\altaffiliation{Participating as a member of TEXONO Collaboration}
\affiliation{Institute of Physics, Academia Sinica, Taipei 11529}
\author{N.~Yi}
\affiliation{Key Laboratory of Particle and Radiation Imaging (Ministry of Education) and Department of Engineering Physics, Tsinghua University, Beijing 100084}
\author{C.~X.~Yu}
\affiliation{School of Physics, Nankai University, Tianjin 300071}
\author{H.~J.~Yu}
\affiliation{NUCTECH Company, Beijing 100084}
\author{J.~F.~Yue}
\affiliation{YaLong River Hydropower Development Company, Chengdu 610051}
\author{M.~Zeng}
\affiliation{Key Laboratory of Particle and Radiation Imaging (Ministry of Education) and Department of Engineering Physics, Tsinghua University, Beijing 100084}
\author{Z.~Zeng}
\affiliation{Key Laboratory of Particle and Radiation Imaging (Ministry of Education) and Department of Engineering Physics, Tsinghua University, Beijing 100084}
\author{B.~T.~Zhang}
\affiliation{Key Laboratory of Particle and Radiation Imaging (Ministry of Education) and Department of Engineering Physics, Tsinghua University, Beijing 100084}
\author {F.~S.~Zhang}
\affiliation{College of Nuclear Science and Technology, Beijing Normal University, Beijing 100875}
\author{L. Zhang}
\affiliation{College of Physics, Sichuan University, Chengdu 610065}
\author{Z.~H.~Zhang}
\affiliation{Key Laboratory of Particle and Radiation Imaging (Ministry of Education) and Department of Engineering Physics, Tsinghua University, Beijing 100084}
\author{K.~K.~Zhao}
\affiliation{College of Physics, Sichuan University, Chengdu 610065}
\author{M.~G.~Zhao}
\affiliation{School of Physics, Nankai University, Tianjin 300071}
\author{J.~F.~Zhou}
\affiliation{YaLong River Hydropower Development Company, Chengdu 610051}
\author{Z.~Y.~Zhou}
\affiliation{Department of Nuclear Physics, China Institute of Atomic Energy, Beijing 102413}
\author{J.~J.~Zhu}
\affiliation{College of Physics, Sichuan University, Chengdu 610065}

\collaboration{CDEX Collaboration}
\noaffiliation

\date{\today}

\begin{abstract}
We present improved germanium-based constraints on sub-GeV dark matter via dark matter--electron ($\chi$-$e$) scattering using the 205.4 kg$\cdot$day dataset from the CDEX-10 experiment. Using a novel calculation technique, we attain predicted $\chi$-$e$ scattering spectra observable in high-purity germanium detectors. In the heavy mediator scenario, our results achieve 3 orders of magnitude of improvement for $m_{\chi}$ larger than 80 MeV/c$^2$ compared to previous germanium-based $\chi$-$e$ results. We also present the most stringent $\chi$-$e$ cross-section limit to date among experiments using solid-state detectors for $m_{\chi}$ larger than 90 MeV/c$^2$ with heavy mediators and $m_{\chi}$ larger than 100 MeV/c$^2$ with electric dipole coupling. The result proves the feasibility and demonstrates the vast potential of a new $\chi$-$e$ detection method with high-purity germanium detectors in ultralow radioactive background.
\end{abstract}

\maketitle

\emph{Introduction.}— 
Current cosmological and astronomical observations strongly indicate the existence of dark matter (DM, denoted as $\chi$) as a major constituent of the Universe~\cite{BERTONE2005279}. Experiments probing DM within the mass range from GeV/$\rm{c}^2$ to TeV/$\rm{c}^2$ via DM-nucleus ($\chi$-$N$) scattering have been widely conducted, such as XENON~\cite{xenon1t}, LUX~\cite{lux}, PandaX~\cite{PandaX}, DarkSide~\cite{darkside}, SuperCDMS~\cite{cdmslite}, and CDEX~\cite{cdex0,cdex1,cdex12014,cdex12016,cdex1b2018,cdex102018,cdex10_tech,c1b2019,cdexmidgal}. Several efforts have been recently made to further extend the experiment reach to lower DM mass ($m_{\chi}$) within the $\chi$-$N$ paradigm by analyzing the physics channel of the Midgal effect, and up until now, a $m_{\chi}$ reach down to $\sim$30 MeV/c$^2$ has been achieved~\cite{luxmigdal,cresstsurface,cdexmidgal,cdexearthshielding}. However, as $m_{\chi}$ further drops, the energy deposited in the detector via $\chi$-$N$ scattering will rapidly decrease to below the thresholds of conventional detection techniques due to mass mismatch between DM particles and heavy nuclei, limiting the probing sensitivity for light DM. This requires us to either further lower the detection threshold or exploit new DM interaction paradigms. The former approach has always been a focus in underground DM experiments, and sufficient efforts have been dedicated to the topic. For the latter approach, the DM-electron ($\chi$-$e$) scattering paradigm proves to be successful among the current DM interaction models. In such a $\chi$-$e$ scattering process, light DM particles can potentially pass most of their energies onto electrons, depositing observable energies onto detectors. Several direct detection experiments have adopted such a paradigm to find the $\chi$-$e$ process, including experiments using solid-state detectors, such as SENSEI~\cite{sensei_chi_e}, DAMIC~\cite{damic_chi_e}, EDELWEISS~\cite{edelweiss_chi_e}, and SuperCDMS~\cite{cdmshvev_chi_e}, and experiments using liquid-noble gases, such as XENON~\cite{xenon10_100_chi_e,xenon1t_chi_e}, PandaX~\cite{pandaxii_chi_e}, and DarkSide~\cite{darkside_chi_e}. These efforts successfully pushed the $m_{\chi}$ reach down to $\sim$1 MeV/c$^2$.

Featuring good energy resolution and ultralow energy threshold~\cite{soma2016}, high-purity germanium (HPGe) detectors have been adopted by CDEX in light DM searches. The CDEX-10 experiment runs a 10-kg p-type point contact HPGe detector array in the China Jinping Underground Laboratory (CJPL) with a rock overburden of 2400 m~\cite{cjpl}. Configuration of the experimental setup is described in detail in Refs.~\cite{cdex102018,cdex10_tech}. With the best performance among the nine detectors in the CDEX-10 experiment, C10B-Ge1 has remained in stable data-taking condition since February 2017, and an analysis threshold of 160 eVee (``eVee" represents electron equivalent energy derived from a charge calibration) has been achieved. The first physical result from CDEX-10, limits on conventional $\chi$-$N$ spin-independent (SI) scattering down to $m_{\chi}$$\sim$2 GeV/c$^2$ are derived with an exposure of 102.8 kg$\cdot$day~\cite{cdex102018}. Subsequently, a larger dataset was acquired in August 2018 with a total live time of 234.6 d. After taking data acquisition (DAQ) dead time into account and necessary event selections to remove nonphysical events caused by electronic noises, the final exposure achieves 205.4 kg$\cdot$day~\cite{cdexdarkphoton,cdexcrdm}, based on which the constraints on dark photon effective mixing parameter were derived~\cite{cdexdarkphoton}. Furthermore, we performed $\chi$-$N$ SI analysis within the paradigm of DM particles boosted by cosmic rays~\cite{cdexcrdm}. With recent advancements in $\chi$-$e$ transition rate calculation techniques for semiconductor detectors, such realms of the $\chi$-$e$ scattering are also unfolded to the CDEX germanium detector. In this Letter, we reanalyzed the 205.4 kg$\cdot$day datasets to set constraints on $\chi$-$e$ interactions.

\emph{Expected rate in Ge detectors.}— 
Via $\chi$-$e$ scattering, DM particles with small $m_{\chi}$ can potentially deposit observable energies onto detectors. Targets, including noble gases with ionization energies of $\mathcal{O}$(10 eV)~\cite{Ionization} and semiconductors with band gaps of $\mathcal{O}$(eV)~\cite{BandGap}, allow such $\chi$-$e$ scattering process to transmit energies up to $\mathcal{O}$(keV). Semiconductor detectors, specifically Si and Ge, characterized by high-energy resolution and low-detection thresholds, are excellent platforms for direct $\chi$-$e$ detection experiments. However, theoretical calculations of $\chi$-$e$ transition rates in semiconductors are much more complicated than in noble gases. In noble gases, atoms can be considered isolated, and wave functions and energy levels are already well tabulated~\cite{AtomWaveFunctions}, whereas atom states are bound with crystal environments for semiconductor targets. Thus, more dedicated calculation techniques are required. By analyzing matrix elements depending only on $\bm{q}$, and assuming electron energy levels to be SI, the $\chi$-$e$ transition rate per target mass $R_{i\rightarrow f}$ is given as follows~\cite{ExceedDM_PRD,chi_e_scattering_rate,light_mediator}:

\begin{align}
R_{i\rightarrow f}&=\frac{2\pi\overline{\sigma}_{e}}{V\mu_{\chi e}^2m_{\chi}}\frac{\rho_{\chi}}{\rho_{T}}\nonumber\\&\cdot\sum_{i,f}\int \frac{d^3q}{(2\pi)^3}\left(\frac{f_e}{f_e^0}\right)^2F^2_{\rm DM}g(\bm{q},\omega)|f_{i\rightarrow f}(\bm{q})|^2,\\
\overline{\sigma}_{e}&=\frac{\mu_{\chi e}^2}{16\pi m_{\chi}^2m_e^2}\overline{|\mathcal{M}(q_0)|^2},\\
f_{i\rightarrow f}&=\int {\rm d}^3{\bm x}e^{i\bm{q}\cdot\bm{x}}\psi^\ast_f(\bm{x})\psi_i(\bm{x}),
\end{align}

where $\rho_{T}$ is the target density, $V$ is the target volume, $\rho_{\chi}$ is the local DM density taken to be 0.3 GeV/cm$^3$~\cite{SHM} following previous similar works~\cite{sensei_chi_e,damic_chi_e,edelweiss_chi_e,cdmshvev_chi_e}, and $\mu_{\chi e}$ is the DM electron reduced mass. $g(\bm{q};\omega)$ is the velocity integral that can be evaluated analytically for the commonly assumed Maxwell-Boltzmann velocity distribution~\cite{ExceedDM_PRD}. In this work, we adopt the standard halo model with the most probable velocity of $v_0=220\rm\ km/s$, the Galactic escape velocity of $v_{\rm esc}=544\rm\ km/s$, and the Earth's velocity of $v_{\rm E}=232\rm\ km/s$ with respect to the Galactic rest frame~\cite{SHM,escapevelocity}. $f_{i\rightarrow f}$ is the momentum transfer dependent crystal form factor; $\overline{\sigma}_{e}$ is the reference cross section for free electron scattering~\cite{chi_e_scattering_rate}. For simple DM models, such as the kinetically mixed dark photon or leptophilic scalar mediator models, spin average matrix element $|\mathcal{M}(\boldsymbol{q})|^2$ can be factorized as $\mathcal{M}(\bm{q})=\mathcal{M}(q_0)(f_e/f_e^0)F_{\rm DM}$. Here, the reference momentum transfer $q_0$ is taken to be $\alpha m_e$; $f_e/f_e^0$ is the screening factor discussed in detail in Ref.~\cite{ExceedDM_PRD}; $F_{\rm DM}$ is the dark matter form factor, where $F_{\rm DM}=1$ corresponds to pointlike interactions with heavy mediators or a magnetic dipole coupling; $F_{\rm DM}=q_0/q$ corresponds to an electric dipole coupling; $F_{\rm DM}=(q_0/q)^2$ corresponds to massless or ultralight mediators.

For $\chi$-$e$ transition rate calculations of semiconductor crystal targets, the tricky part is the crystal form factor calculation. Several attempts have been dedicated to this topic, including semianalytic approximations in Refs.~\cite{BandGap,SemiAnalytic1} and a fully numerical approach that employs density-functional theory (DFT) in Refs.~\cite{Ionization,QEDark}, which is currently the standard first-principle calculation method of $\chi$-$e$ transition rates and has been adopted by several semiconductor-based experiments~\cite{sensei_chi_e,damic_chi_e,edelweiss_chi_e,cdmshvev_chi_e}. Recently, the relationship between the dielectric function and the SI scattering rate has been investigated~\cite{dielectric_scattering}, and a calculation method using energy loss function (ELF) has been proposed~\cite{elf,darkelf}. A possibility of using the polarization function to compute the scattering rate through scalar interactions has been explored as well~\cite{scalarinteraction}. Combined with advanced ultralow threshold detector technologies, these methods have gained great success in the $\chi$-$e$ scattering probing. Nevertheless, since these methods focus more on the low-energy region within a few tens of eV that contributes the majority of the scattering rates, which is lower than the typical HPGe detectors' threshold of $\mathcal{O}$(100 eV), we are substantially prevented from using these methods to probe $\chi$-$e$ scattering in our CDEX experiments.

However, with another novel calculation technique combining DFT and semianalytic methods presented by Ref.~\cite{ExceedDM_PRD}, predicted spectra can be expanded to $\mathcal{O}$(keV), opening a new channel for conventional HPGe detector-based experiments to probe light DM particles via $\chi$-$e$ scattering. The newly developed method improves in two key aspects compared to previous works: implementing all-electron (AE) reconstruction to recover high momentum components in wave functions calculated using DFT~\cite{AE} and extending the calculation to bands further away from the band gap using semianalytic approximations. In such a calculation process, crystal electronic states are divided into four categories: core, valence, conduction, and free, and the Fermi energy, defined as the top of the valence bands, is denoted by $E=0$. For Ge, the first four bands below the band gap are treated as valence, and its energy spans from 0 to $-$14 eV. Additionally, bands up to $E=60$ eV are treated as conduction. These highly perturbed states are computed using DFT methods. Meanwhile, electrons in bands below $-$14 eV ($1s$ to $3d$) and above 60 eV are more isolated from the crystal environment, making it accurate to approximate them to core electrons and plane waves; these bands are denoted by core and free. For plane-wave approximation, an effective charge parameter $Z_{\rm eff}$ is taken to be 1~\cite{ExceedDM_PRD}.  With electron states in Ge crystal fully modeled, a more complete prediction of $\chi$-$e$ scattering rate is attained. This novel calculation technique is implemented in the $\tt EXCEED$-$\tt DM$ package~\cite{ExceedDM020}. Similar work is done via effective field theory in Ref.~\cite{effectivefieldtheory}.

\begin{figure}[!htbp]
\includegraphics[width=\linewidth]{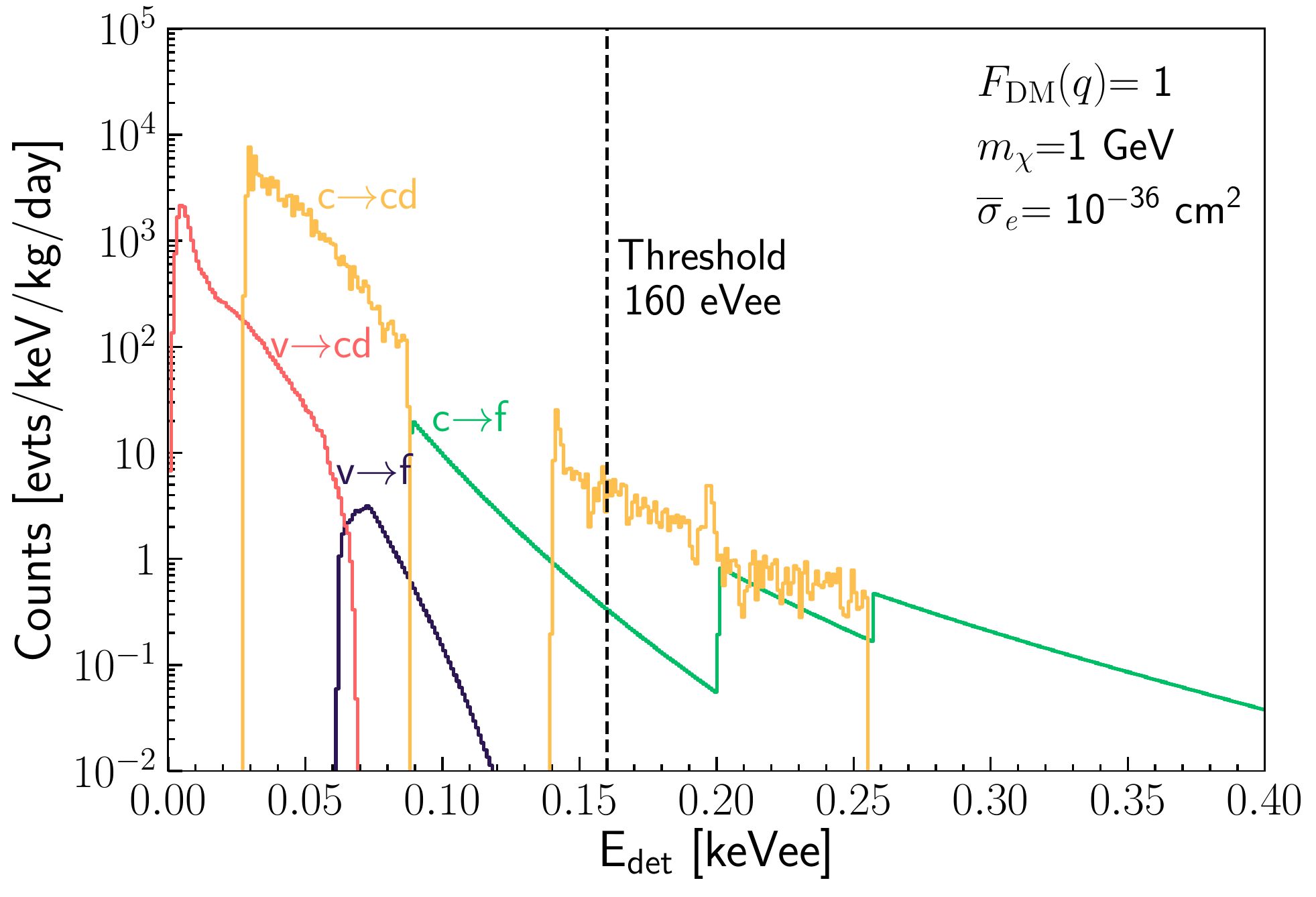}
\caption{
Predicted scattering rates calculated by $\tt EXCEED$-$\tt DM$ of 1 GeV/c$^2$ DM particles on Ge targets for the heavy mediator scenario, with $\overline{\sigma}_{e}=10^{-36}\rm\ cm^2$. Contributions from v$\rightarrow$cd, v$\rightarrow$f, c$\rightarrow$cd, c$\rightarrow$f transitions are depicted. The energy resolution is not considered in this plot. The black dashed line represents the energy threshold of C10B-Ge1.
} 
\label{fig::Components}
\end{figure}

Figure~\ref{fig::Components} shows the calculated contributions from four transition types: valence to conduction (v$\rightarrow$cd), valence to free (v$\rightarrow$f), core to conduction (c$\rightarrow$cd), and core to free (c$\rightarrow$f). The total rate is the sum of the four types of contributions. Compared with previous works~\cite{QEDark}, the new technique expands predicted spectra well above HPGe detectors' thresholds, where the major contributions come from previously ignored c$\rightarrow$cd and c$\rightarrow$f transitions, which enables us to perform $\chi$-$e$ scattering analysis on HPGe detectors.

\begin{figure}[!htbp]
\includegraphics[width=\linewidth]{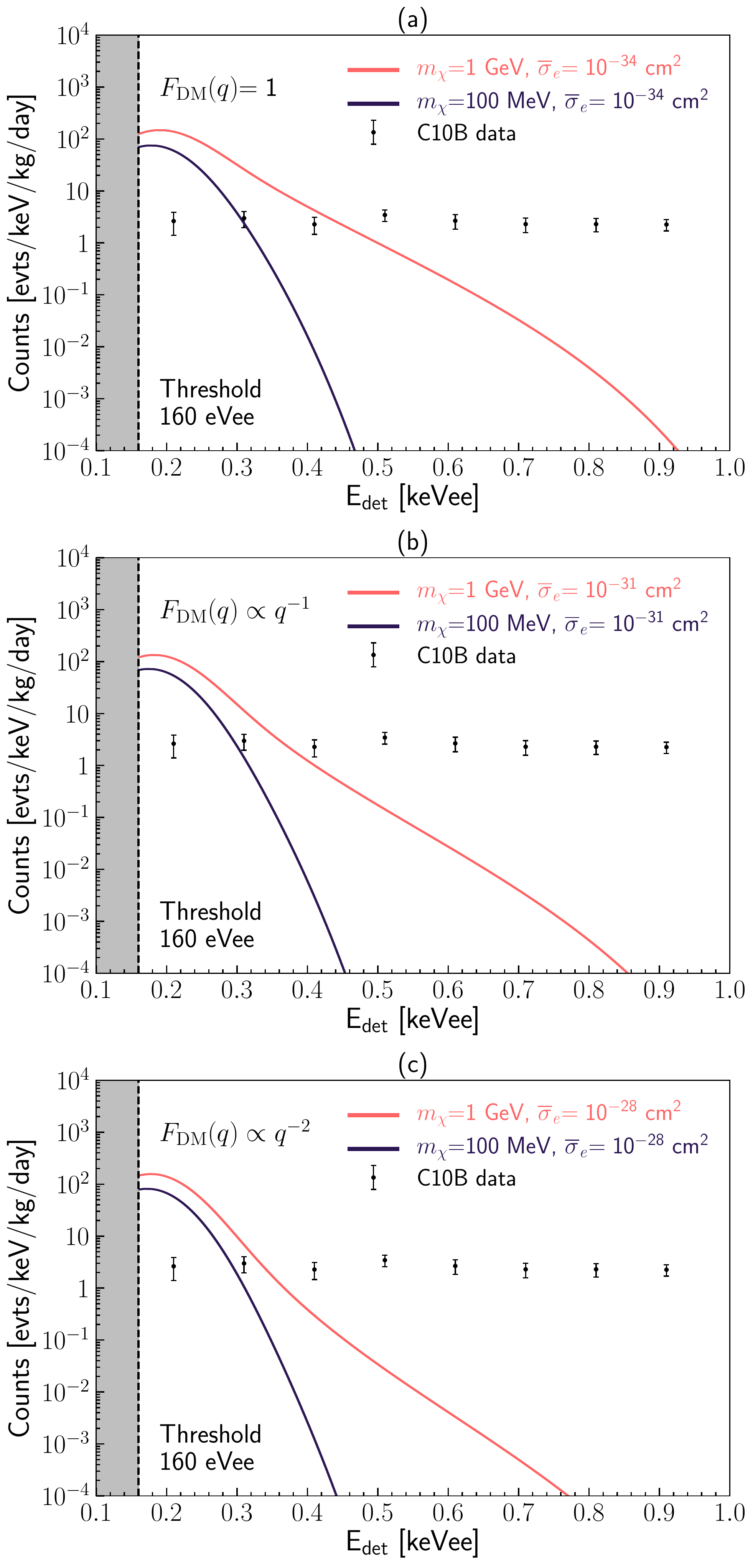}
\caption{
(a) Calculated spectra of 1 GeV/c$^2$ and 100 MeV/c$^2$ DM particles with $\overline{\sigma}_{e}=10^{-34}\rm\ cm^2$ for the heavy mediator scenario. The energy resolution is considered in this plot, and its standard deviation is determined by $\sigma = 35.8 + 16.6\times\sqrt{E}\rm\ (eV)$~\cite{cdexcrdm}, where $E$ is in keV. The contributions from predicted spectrum below 160 eVee (shaded area) are removed before convolving with energy resolution. The black points with error bars represents the measured spectrum from C10B-Ge1 with a 205.4 kg$\cdot$day exposure. The energy range and bin width are 0.16$-$2.16 keVee and 100 eVee, respectively. (b) Calculated spectra of 1 GeV/c$^2$ and 100 MeV/c$^2$ DM particles with $\overline{\sigma}_{e}=10^{-31}\rm\ cm^2$ for electric dipole coupling. (c)  Calculated spectra of 1 GeV/c$^2$ and 100 MeV/c$^2$ DM particles with $\overline{\sigma}_{e}=10^{-28}\rm\ cm^2$ for the light mediator scenario.
} 
\label{fig::GeSpectrum}
\end{figure}

Figure~\ref{fig::GeSpectrum} shows the total spectra convolved with energy resolution for different $F_{\rm{DM}}$. Since the effect of energy resolution on small energy events has not yet been tested by experiments, an energy resolution effect boundary of 160 eVee (same as the analysis threshold, represented by shaded area in Fig.~\ref{fig::GeSpectrum}) is adopted, below which the contribution from the predicted spectrum is removed before convolving with energy resolution.

\emph{Data analysis.}— 
Data analysis of this work is based on a 205.4 kg$\cdot$day dataset from C10B-Ge1, and follows the procedures established in previous works~\cite{cdex12016,cdex102018,cdex1b2018}. Energy calibrations were done by the zero energy (defined by the random trigger events) and the internal cosmogenic K-shell x-ray peaks: 8.98 keVee of $\rm ^{65}Zn$ and 10.37 keVee of $\rm ^{68,71}Ge$. The signal events are identified after pedestal noise cut, physics event selection, and bulk or surface event discrimination~\cite{Li:2014a,Yang:2018a}. The detailed procedures and corresponding efficiencies of this dataset can be found in Refs.~\cite{cdex102018,cdex10_tech,cdexdarkphoton}. The physics analysis threshold is set to be 160 eVee where the combined efficiency is 4.5\%. At the sub-keVee energy range relevant to this analysis, background events are dominated by Compton scattering of high-energy gamma rays and internal radioactivity from long-lived cosmogenic isotopes. Figure~\ref{fig::GeSpectrum} shows the measured spectrum after subtracting the contributions from L- and M-shell x-ray peaks derived from the corresponding K-shell line intensities~\cite{cdex102018,cdexdarkphoton,cdexcrdm}.

A minimum-$\chi^2$ analysis~\cite{cdex12014} is applied to the residual spectrum at the range of 0.16$-$2.16 keVee:
\begin{equation}
\begin{aligned}
\label{con:chi2}
\chi^2(m_{\chi},\overline{\sigma}_{e})=\sum_{i=1}^N\frac{[n_i-B-S_i(m_{\chi},\overline{\sigma}_{e})]^2}{\Delta_i^2},
\end{aligned}
\end{equation}
where $n_i$ and $\Delta_i$ are measured data and standard deviation with statistical and systematical components at the $i$th energy bin, respectively; $S_i(m_{\chi},\overline{\sigma}_{e})$ is the $\chi$-$e$ scattering rate prediction; $B$ is the assumed flat background contribution from the Compton scattering of high-energy gamma rays.

\begin{figure}[!htbp]
\includegraphics[width=\linewidth]{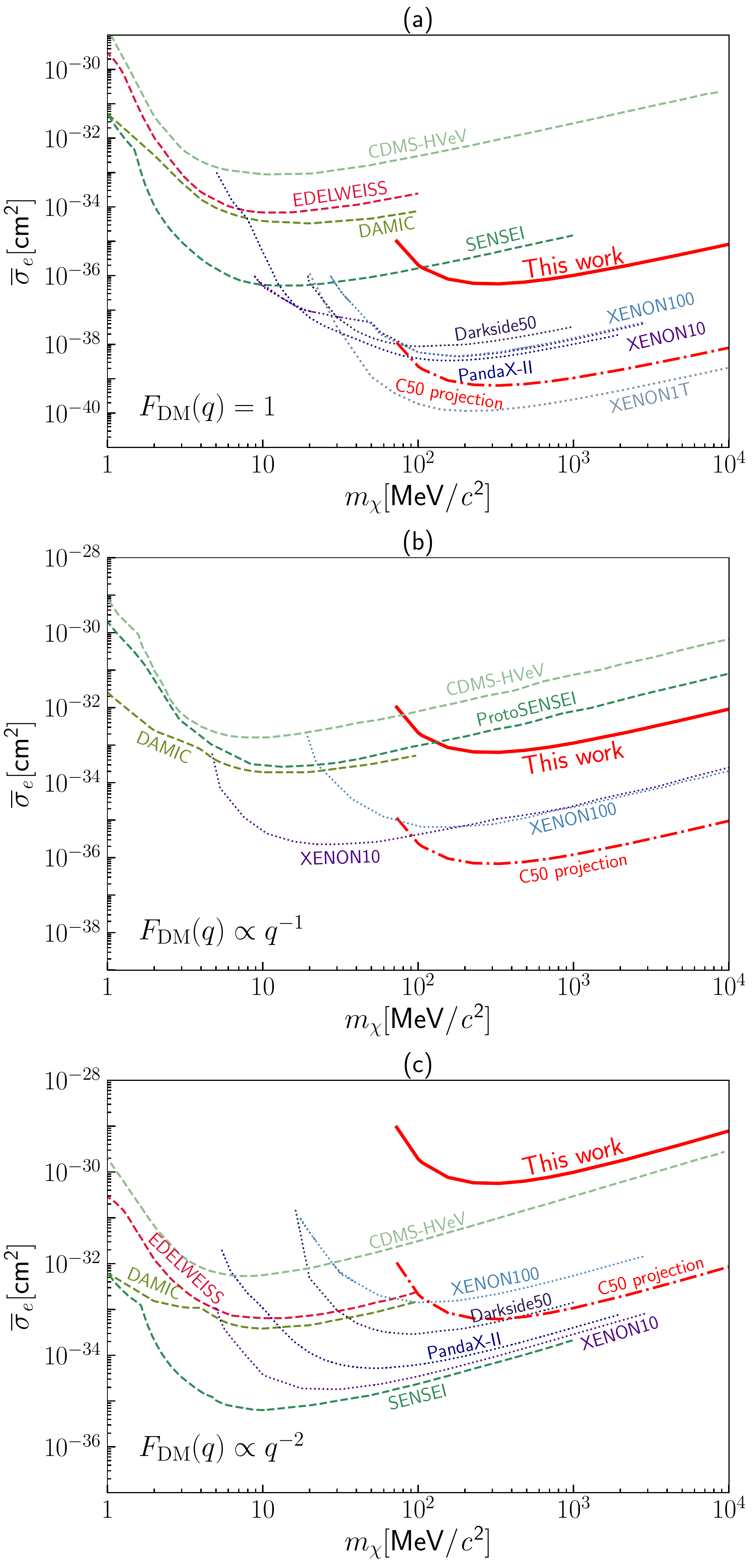}
\caption{
90\% C.L. upper limits on the DM-electron free scattering cross section $\overline{\sigma}_{e}$  as a function of DM mass $m_{\chi}$ for $F_{\rm DM}\propto q^{-n} (n=0,1,2)$ from CDEX-10 by superimposing the results from several solid-state detector-based~\cite{sensei_chi_e,damic_chi_e,edelweiss_chi_e,cdmshvev_chi_e} (dashed line) and liquid noble gas-based~\cite{xenon10_100_chi_e,xenon1t_chi_e,pandaxii_chi_e,darkside_chi_e} (dotted line) experiments, and projections of the future CDEX-50 experiment (gray dash-dotted line). Constraints and projections proposed here correspond to an energy threshold of 160 eVee and the same boundary of energy resolution effect. The only germanium-based $\chi$-$e$ results presented before this work from EDELWEISS are marked with crimson dashed lines. (a) Heavy mediator scenario. We present the strongest limit among solid-state detector-based experiments for $m_{\chi}>90$ MeV/c$^2$. (b) Electric dipole coupling scenario. Analyses on protoSENSEI at MINOS~\cite{protosensei_chi_e}, CDMS-HVeV surface run~\cite{cdmshvev_chi_e_r1}, XENON10, and XENON100~\cite{xenon10_100_chi_e} are presented by Emken $et\ al.$ in Ref.~\cite{earthshielding_chi_e}. (c) Light mediator scenario.
}
\label{fig::ExclusionLine}
\end{figure}

With predicted spectra calculated by $\tt EXCEED$-$\tt DM$, a 90$\%$ confidence level (C.L.) one-side upper limit exclusion line of $\overline{\sigma}_{e}$ with $\Delta\chi^2=1.64$ is derived~\cite{chisquare}. Figures~\ref{fig::ExclusionLine}(a)--(c) show the results for heavy mediators, electric dipole coupling, and light mediators, respectively. It is worth noting that the Earth shielding effects are negligible at the level of excluded cross sections~\cite{earthshielding_chi_e}. Results from several liquid noble gas-based~\cite{xenon10_100_chi_e,xenon1t_chi_e,pandaxii_chi_e,darkside_chi_e} and solid-state detector-based experiments analyzed with $\tt{QEDARK}$~\cite{sensei_chi_e,damic_chi_e,edelweiss_chi_e,cdmshvev_chi_e} are also depicted in Fig.~\ref{fig::ExclusionLine}.

In principle, due to upward fluctuations from energy resolution smearing ~\cite{cdex10_tech}, events with energy depositions lower than the analysis threshold can still have finite probability to satisfy the trigger conditions and survive the signal selection criteria. However, the signal response model of these events is not yet completely established; hence, we adopt a conservative approach in this analysis to derive sensitivity constraints that only $\chi$-$e$ events predicted by $\tt EXCEED$-$\tt DM$ with raw recoil energy above 160 eVee (same as the analysis threshold applied to measured spectra) are considered. Response of events with subthreshold energy depositions will be studied in our future research, and contributions to the $\chi$-$e$ sensitivity from this kind of event will be further evaluated by then.

Compared with previous germanium-based results from EDELWEISS using the cryogenic calorimeter technique, our results achieve 3 orders of magnitude of improvement for $m_{\chi}$ larger than 80 MeV/c$^2$. As illustrated in Fig.~\ref{fig::ExclusionLine}(a), in the heavy mediator scenario, our result proves to be most stringent among solid-state detector-based experiments in high mass region of $m_{\chi}$ $>$ 90 MeV/c$^2$. As shown in Fig.~\ref{fig::ExclusionLine}(b) with electric dipole coupling, our result also takes the lead for $m_{\chi}$ larger than 100 MeV/$c^2$ among solid state experiment results.

\emph{Summary.}— 
In this Letter, we exploit a new route of $\chi$-$e$ scattering probing through high-purity germanium detector technology and present the first HPGe detector-based $\chi$-$e$ scattering limits from the CDEX experiment, significantly surpassing previous germanium bolometer-based results from EDELWEISS for $m_{\chi}$ $>$ 80 MeV/c$^2$. For the heavy mediator scenario and the electric dipole coupling scenario, we present leading constraints on $\chi$-$e$ scattering cross sections among solid-state detector-based experiments in high mass range. These results reveal the vast potential of such technical route in the realm of $\chi$-$e$ scattering probing, paving the path for our journey to the frontier of sub-GeV dark matter search.

For germanium detectors, the expected $\chi$-$e$ scattering rate drops drastically as deposited energy increases. Additionally, compared to other solid-state detector-based experiments~\cite{sensei_chi_e,damic_chi_e,edelweiss_chi_e,cdmshvev_chi_e}, the energy thresholds of HPGe detectors are not ideal, hindering us in our sensitivity to $\chi$-$e$ scattering. However, the superior ultralow radiation environment and significantly larger exposure of the CDEX experiment greatly compensate for this loss in sensitivity, helping us achieve a competitive probing ability for the $\chi$-$e$ interactions. Detection sensitivity of $\chi$-$e$ scattering for future CDEX experiments will be further enhanced with lower radioactive backgrounds and energy thresholds.

CDEX-50, the next generation of the CDEX experiment, is currently in preparation. CDEX-50 will use an array of 50 1-kg HPGe detectors with optimized electronics and will be operated in a superior radioactive environment. For the CDEX-50 experiment, the radioactive background will be further reduced to $\sim$0.01 evts/keV/kg/day in the sub-keV region, i.e., 200 times lower than our current background level. As shown in Fig.~\ref{fig::ExclusionLine}, with improved radioactive background, exposure of 50 kg$\cdot$y and energy threshold of 160 eVee, it is anticipated that an improvement up to 3 orders of magnitude can be achieved compared to our current $\overline{\sigma}_{e}$ limits.
 
We thank Tanner Trickle for his help on the $\tt EXCEED$-$\tt DM$ package and Yufeng Zhou for his insightful comments. This work was supported by the National Key Research and Development Program of China (Grant No. 2017YFA0402200) and the National Natural Science Foundation of China (Grants No. 12175112, No. 12005111, and No. 11725522). We acknowledge the Center of High-Performance Computing, Tsinghua University, for providing the facility support.

\bibliography{ElectronRecoil.bib}

 \newcommand{\noop}[1]{}
\begin{thebibliography}{53}%
\makeatletter
\providecommand \@ifxundefined [1]{%
 \@ifx{#1\undefined}
}%
\providecommand \@ifnum [1]{%
 \ifnum #1\expandafter \@firstoftwo
 \else \expandafter \@secondoftwo
 \fi
}%
\providecommand \@ifx [1]{%
 \ifx #1\expandafter \@firstoftwo
 \else \expandafter \@secondoftwo
 \fi
}%
\providecommand \natexlab [1]{#1}%
\providecommand \enquote  [1]{``#1''}%
\providecommand \bibnamefont  [1]{#1}%
\providecommand \bibfnamefont [1]{#1}%
\providecommand \citenamefont [1]{#1}%
\providecommand \href@noop [0]{\@secondoftwo}%
\providecommand \href [0]{\begingroup \@sanitize@url \@href}%
\providecommand \@href[1]{\@@startlink{#1}\@@href}%
\providecommand \@@href[1]{\endgroup#1\@@endlink}%
\providecommand \@sanitize@url [0]{\catcode `\\12\catcode `\$12\catcode
  `\&12\catcode `\#12\catcode `\^12\catcode `\_12\catcode `\%12\relax}%
\providecommand \@@startlink[1]{}%
\providecommand \@@endlink[0]{}%
\providecommand \url  [0]{\begingroup\@sanitize@url \@url }%
\providecommand \@url [1]{\endgroup\@href {#1}{\urlprefix }}%
\providecommand \urlprefix  [0]{URL }%
\providecommand \Eprint [0]{\href }%
\providecommand \doibase [0]{http://dx.doi.org/}%
\providecommand \selectlanguage [0]{\@gobble}%
\providecommand \bibinfo  [0]{\@secondoftwo}%
\providecommand \bibfield  [0]{\@secondoftwo}%
\providecommand \translation [1]{[#1]}%
\providecommand \BibitemOpen [0]{}%
\providecommand \bibitemStop [0]{}%
\providecommand \bibitemNoStop [0]{.\EOS\space}%
\providecommand \EOS [0]{\spacefactor3000\relax}%
\providecommand \BibitemShut  [1]{\csname bibitem#1\endcsname}%
\let\auto@bib@innerbib\@empty
\bibitem [{\citenamefont {Bertone}\ \emph {et~al.}(2005)\citenamefont
  {Bertone}, \citenamefont {Hooper},\ and\ \citenamefont
  {Silk}}]{BERTONE2005279}%
  \BibitemOpen
  \bibfield  {author} {\bibinfo {author} {\bibfnamefont {G.}~\bibnamefont
  {Bertone}}, \bibinfo {author} {\bibfnamefont {D.}~\bibnamefont {Hooper}}, \
  and\ \bibinfo {author} {\bibfnamefont {J.}~\bibnamefont {Silk}},\ }\href
  {\doibase https://doi.org/10.1016/j.physrep.2004.08.031} {\bibfield
  {journal} {\bibinfo  {journal} {Phys. Rep.}\ }\textbf {\bibinfo {volume}
  {405}},\ \bibinfo {pages} {279} (\bibinfo {year} {2005})}\BibitemShut
  {NoStop}%
\bibitem [{\citenamefont {Aprile}\ \emph {et~al.}(2018)\citenamefont {Aprile}
  \emph {et~al.}}]{xenon1t}%
  \BibitemOpen
  \bibfield  {author} {\bibinfo {author} {\bibfnamefont {E.}~\bibnamefont
  {Aprile}} \emph {et~al.} (\bibinfo {collaboration} {XENON Collaboration}),\
  }\href {\doibase 10.1103/PhysRevLett.121.111302} {\bibfield  {journal}
  {\bibinfo  {journal} {Phys. Rev. Lett.}\ }\textbf {\bibinfo {volume} {121}},\
  \bibinfo {pages} {111302} (\bibinfo {year} {2018})}\BibitemShut {NoStop}%
\bibitem [{\citenamefont {Akerib}\ \emph {et~al.}(2017)\citenamefont {Akerib}
  \emph {et~al.}}]{lux}%
  \BibitemOpen
  \bibfield  {author} {\bibinfo {author} {\bibfnamefont {D.~S.}\ \bibnamefont
  {Akerib}} \emph {et~al.} (\bibinfo {collaboration} {LUX Collaboration}),\
  }\href {\doibase 10.1103/PhysRevLett.118.021303} {\bibfield  {journal}
  {\bibinfo  {journal} {Phys. Rev. Lett.}\ }\textbf {\bibinfo {volume} {118}},\
  \bibinfo {pages} {021303} (\bibinfo {year} {2017})}\BibitemShut {NoStop}%
\bibitem [{\citenamefont {Cui}\ \emph {et~al.}(2017)\citenamefont {Cui} \emph
  {et~al.}}]{PandaX}%
  \BibitemOpen
  \bibfield  {author} {\bibinfo {author} {\bibfnamefont {X.}~\bibnamefont
  {Cui}} \emph {et~al.} (\bibinfo {collaboration} {PandaX-II Collaboration}),\
  }\href {\doibase 10.1103/PhysRevLett.119.181302} {\bibfield  {journal}
  {\bibinfo  {journal} {Phys. Rev. Lett.}\ }\textbf {\bibinfo {volume} {119}},\
  \bibinfo {pages} {181302} (\bibinfo {year} {2017})}\BibitemShut {NoStop}%
\bibitem [{\citenamefont {Agnes}\ \emph
  {et~al.}(2018{\natexlab{a}})\citenamefont {Agnes} \emph {et~al.}}]{darkside}%
  \BibitemOpen
  \bibfield  {author} {\bibinfo {author} {\bibfnamefont {P.}~\bibnamefont
  {Agnes}} \emph {et~al.} (\bibinfo {collaboration} {DarkSide Collaboration}),\
  }\href {\doibase 10.1103/PhysRevLett.121.081307} {\bibfield  {journal}
  {\bibinfo  {journal} {Phys. Rev. Lett.}\ }\textbf {\bibinfo {volume} {121}},\
  \bibinfo {pages} {081307} (\bibinfo {year} {2018}{\natexlab{a}})}\BibitemShut
  {NoStop}%
\bibitem [{\citenamefont {Agnese}\ \emph
  {et~al.}(2018{\natexlab{a}})\citenamefont {Agnese} \emph
  {et~al.}}]{cdmslite}%
  \BibitemOpen
  \bibfield  {author} {\bibinfo {author} {\bibfnamefont {R.}~\bibnamefont
  {Agnese}} \emph {et~al.} (\bibinfo {collaboration} {SuperCDMS
  Collaboration}),\ }\href {\doibase 10.1103/PhysRevD.97.022002} {\bibfield
  {journal} {\bibinfo  {journal} {Phys. Rev. D}\ }\textbf {\bibinfo {volume}
  {97}},\ \bibinfo {pages} {022002} (\bibinfo {year}
  {2018}{\natexlab{a}})}\BibitemShut {NoStop}%
\bibitem [{\citenamefont {Liu}\ \emph {et~al.}(2014)\citenamefont {Liu} \emph
  {et~al.}}]{cdex0}%
  \BibitemOpen
  \bibfield  {author} {\bibinfo {author} {\bibfnamefont {S.~K.}\ \bibnamefont
  {Liu}} \emph {et~al.} (\bibinfo {collaboration} {CDEX Collaboration}),\
  }\href {\doibase 10.1103/PhysRevD.90.032003} {\bibfield  {journal} {\bibinfo
  {journal} {Phys. Rev. D}\ }\textbf {\bibinfo {volume} {90}},\ \bibinfo
  {pages} {032003} (\bibinfo {year} {2014})}\BibitemShut {NoStop}%
\bibitem [{\citenamefont {Zhao}\ \emph {et~al.}(2013)\citenamefont {Zhao} \emph
  {et~al.}}]{cdex1}%
  \BibitemOpen
  \bibfield  {author} {\bibinfo {author} {\bibfnamefont {W.}~\bibnamefont
  {Zhao}} \emph {et~al.} (\bibinfo {collaboration} {CDEX Collaboration}),\
  }\href {\doibase 10.1103/PhysRevD.88.052004} {\bibfield  {journal} {\bibinfo
  {journal} {Phys. Rev. D}\ }\textbf {\bibinfo {volume} {88}},\ \bibinfo
  {pages} {052004} (\bibinfo {year} {2013})}\BibitemShut {NoStop}%
\bibitem [{\citenamefont {Yue}\ \emph {et~al.}(2014)\citenamefont {Yue} \emph
  {et~al.}}]{cdex12014}%
  \BibitemOpen
  \bibfield  {author} {\bibinfo {author} {\bibfnamefont {Q.}~\bibnamefont
  {Yue}} \emph {et~al.} (\bibinfo {collaboration} {CDEX Collaboration}),\
  }\href {\doibase 10.1103/PhysRevD.90.091701} {\bibfield  {journal} {\bibinfo
  {journal} {Phys. Rev. D}\ }\textbf {\bibinfo {volume} {90}},\ \bibinfo
  {pages} {091701} (\bibinfo {year} {2014})}\BibitemShut {NoStop}%
\bibitem [{\citenamefont {Zhao}\ \emph {et~al.}(2016)\citenamefont {Zhao} \emph
  {et~al.}}]{cdex12016}%
  \BibitemOpen
  \bibfield  {author} {\bibinfo {author} {\bibfnamefont {W.}~\bibnamefont
  {Zhao}} \emph {et~al.} (\bibinfo {collaboration} {CDEX Collaboration}),\
  }\href {\doibase 10.1103/PhysRevD.93.092003} {\bibfield  {journal} {\bibinfo
  {journal} {Phys. Rev. D}\ }\textbf {\bibinfo {volume} {93}},\ \bibinfo
  {pages} {092003} (\bibinfo {year} {2016})}\BibitemShut {NoStop}%
\bibitem [{\citenamefont {Yang}\ \emph
  {et~al.}(2018{\natexlab{a}})\citenamefont {Yang} \emph
  {et~al.}}]{cdex1b2018}%
  \BibitemOpen
  \bibfield  {author} {\bibinfo {author} {\bibfnamefont {L.~T.}\ \bibnamefont
  {Yang}} \emph {et~al.} (\bibinfo {collaboration} {CDEX Collaboration}),\
  }\href {\doibase 10.1088/1674-1137/42/2/023002} {\bibfield  {journal}
  {\bibinfo  {journal} {Chin. Phys. C}\ }\textbf {\bibinfo {volume} {42}},\
  \bibinfo {eid} {023002} (\bibinfo {year} {2018}{\natexlab{a}})}\BibitemShut
  {NoStop}%
\bibitem [{\citenamefont {Jiang}\ \emph {et~al.}(2018)\citenamefont {Jiang}
  \emph {et~al.}}]{cdex102018}%
  \BibitemOpen
  \bibfield  {author} {\bibinfo {author} {\bibfnamefont {H.}~\bibnamefont
  {Jiang}} \emph {et~al.} (\bibinfo {collaboration} {CDEX Collaboration}),\
  }\href {\doibase 10.1103/PhysRevLett.120.241301} {\bibfield  {journal}
  {\bibinfo  {journal} {Phys. Rev. Lett.}\ }\textbf {\bibinfo {volume} {120}},\
  \bibinfo {pages} {241301} (\bibinfo {year} {2018})}\BibitemShut {NoStop}%
\bibitem [{\citenamefont {Jiang}\ \emph {et~al.}(2019)\citenamefont {Jiang}
  \emph {et~al.}}]{cdex10_tech}%
  \BibitemOpen
  \bibfield  {author} {\bibinfo {author} {\bibfnamefont {H.}~\bibnamefont
  {Jiang}} \emph {et~al.} (\bibinfo {collaboration} {CDEX Collaboration}),\
  }\href {\doibase 10.1007/s11433-018-8001-3} {\bibfield  {journal} {\bibinfo
  {journal} {Sci. China Phys. Mech. Astron.}\ }\textbf {\bibinfo {volume}
  {62}},\ \bibinfo {pages} {031012} (\bibinfo {year} {2019})}\BibitemShut
  {NoStop}%
\bibitem [{\citenamefont {Yang}\ \emph {et~al.}(2019)\citenamefont {Yang} \emph
  {et~al.}}]{c1b2019}%
  \BibitemOpen
  \bibfield  {author} {\bibinfo {author} {\bibfnamefont {L.~T.}\ \bibnamefont
  {Yang}} \emph {et~al.} (\bibinfo {collaboration} {CDEX Collaboration}),\
  }\href {\doibase 10.1103/PhysRevLett.123.221301} {\bibfield  {journal}
  {\bibinfo  {journal} {Phys. Rev. Lett.}\ }\textbf {\bibinfo {volume} {123}},\
  \bibinfo {pages} {221301} (\bibinfo {year} {2019})}\BibitemShut {NoStop}%
\bibitem [{\citenamefont {Liu}\ \emph {et~al.}(2019)\citenamefont {Liu} \emph
  {et~al.}}]{cdexmidgal}%
  \BibitemOpen
  \bibfield  {author} {\bibinfo {author} {\bibfnamefont {Z.~Z.}\ \bibnamefont
  {Liu}} \emph {et~al.} (\bibinfo {collaboration} {CDEX Collaboration}),\
  }\href {\doibase 10.1103/PhysRevLett.123.161301} {\bibfield  {journal}
  {\bibinfo  {journal} {Phys. Rev. Lett.}\ }\textbf {\bibinfo {volume} {123}},\
  \bibinfo {pages} {161301} (\bibinfo {year} {2019})}\BibitemShut {NoStop}%
\bibitem [{\citenamefont {Akerib}\ \emph {et~al.}(2019)\citenamefont {Akerib}
  \emph {et~al.}}]{luxmigdal}%
  \BibitemOpen
  \bibfield  {author} {\bibinfo {author} {\bibfnamefont {D.~S.}\ \bibnamefont
  {Akerib}} \emph {et~al.} (\bibinfo {collaboration} {LUX Collaboration}),\
  }\href {\doibase 10.1103/PhysRevLett.122.131301} {\bibfield  {journal}
  {\bibinfo  {journal} {Phys. Rev. Lett.}\ }\textbf {\bibinfo {volume} {122}},\
  \bibinfo {pages} {131301} (\bibinfo {year} {2019})}\BibitemShut {NoStop}%
\bibitem [{\citenamefont {Angloher}\ \emph {et~al.}(2017)\citenamefont
  {Angloher} \emph {et~al.}}]{cresstsurface}%
  \BibitemOpen
  \bibfield  {author} {\bibinfo {author} {\bibfnamefont {G.}~\bibnamefont
  {Angloher}} \emph {et~al.} (\bibinfo {collaboration} {CRESST
  Collaboration}),\ }\href {\doibase 10.1140/epjc/s10052-017-5223-9} {\bibfield
   {journal} {\bibinfo  {journal} {Eur. Phys. J. C}\ }\textbf {\bibinfo
  {volume} {77}},\ \bibinfo {pages} {637} (\bibinfo {year} {2017})}\BibitemShut
  {NoStop}%
\bibitem [{\citenamefont {Liu}\ \emph {et~al.}(2022)\citenamefont {Liu} \emph
  {et~al.}}]{cdexearthshielding}%
  \BibitemOpen
  \bibfield  {author} {\bibinfo {author} {\bibfnamefont {Z.~Z.}\ \bibnamefont
  {Liu}} \emph {et~al.} (\bibinfo {collaboration} {CDEX Collaboration}),\
  }\href {\doibase 10.1103/PhysRevD.105.052005} {\bibfield  {journal} {\bibinfo
   {journal} {Phys. Rev. D}\ }\textbf {\bibinfo {volume} {105}},\ \bibinfo
  {pages} {052005} (\bibinfo {year} {2022})}\BibitemShut {NoStop}%
\bibitem [{\citenamefont {Barak}\ \emph {et~al.}(2020)\citenamefont {Barak}
  \emph {et~al.}}]{sensei_chi_e}%
  \BibitemOpen
  \bibfield  {author} {\bibinfo {author} {\bibfnamefont {L.}~\bibnamefont
  {Barak}} \emph {et~al.} (\bibinfo {collaboration} {SENSEI Collaboration}),\
  }\href {\doibase 10.1103/PhysRevLett.125.171802} {\bibfield  {journal}
  {\bibinfo  {journal} {Phys. Rev. Lett.}\ }\textbf {\bibinfo {volume} {125}},\
  \bibinfo {pages} {171802} (\bibinfo {year} {2020})}\BibitemShut {NoStop}%
\bibitem [{\citenamefont {Aguilar-Arevalo}\ \emph {et~al.}(2019)\citenamefont
  {Aguilar-Arevalo} \emph {et~al.}}]{damic_chi_e}%
  \BibitemOpen
  \bibfield  {author} {\bibinfo {author} {\bibfnamefont {A.}~\bibnamefont
  {Aguilar-Arevalo}} \emph {et~al.} (\bibinfo {collaboration} {DAMIC
  Collaboration}),\ }\href {\doibase 10.1103/PhysRevLett.123.181802} {\bibfield
   {journal} {\bibinfo  {journal} {Phys. Rev. Lett.}\ }\textbf {\bibinfo
  {volume} {123}},\ \bibinfo {pages} {181802} (\bibinfo {year}
  {2019})}\BibitemShut {NoStop}%
\bibitem [{\citenamefont {Arnaud}\ \emph {et~al.}(2020)\citenamefont {Arnaud}
  \emph {et~al.}}]{edelweiss_chi_e}%
  \BibitemOpen
  \bibfield  {author} {\bibinfo {author} {\bibfnamefont {Q.}~\bibnamefont
  {Arnaud}} \emph {et~al.} (\bibinfo {collaboration} {EDELWEISS
  Collaboration}),\ }\href {\doibase 10.1103/PhysRevLett.125.141301} {\bibfield
   {journal} {\bibinfo  {journal} {Phys. Rev. Lett.}\ }\textbf {\bibinfo
  {volume} {125}},\ \bibinfo {pages} {141301} (\bibinfo {year}
  {2020})}\BibitemShut {NoStop}%
\bibitem [{\citenamefont {Amaral}\ \emph {et~al.}(2020)\citenamefont {Amaral}
  \emph {et~al.}}]{cdmshvev_chi_e}%
  \BibitemOpen
  \bibfield  {author} {\bibinfo {author} {\bibfnamefont {D.~W.}\ \bibnamefont
  {Amaral}} \emph {et~al.} (\bibinfo {collaboration} {SuperCDMS
  Collaboration}),\ }\href {\doibase 10.1103/PhysRevD.102.091101} {\bibfield
  {journal} {\bibinfo  {journal} {Phys. Rev. D}\ }\textbf {\bibinfo {volume}
  {102}},\ \bibinfo {pages} {091101} (\bibinfo {year} {2020})}\BibitemShut
  {NoStop}%
\bibitem [{\citenamefont {Essig}\ \emph {et~al.}(2017)\citenamefont {Essig},
  \citenamefont {Volansky},\ and\ \citenamefont {Yu}}]{xenon10_100_chi_e}%
  \BibitemOpen
  \bibfield  {author} {\bibinfo {author} {\bibfnamefont {R.}~\bibnamefont
  {Essig}}, \bibinfo {author} {\bibfnamefont {T.}~\bibnamefont {Volansky}}, \
  and\ \bibinfo {author} {\bibfnamefont {T.-T.}\ \bibnamefont {Yu}},\ }\href
  {\doibase 10.1103/PhysRevD.96.043017} {\bibfield  {journal} {\bibinfo
  {journal} {Phys. Rev. D}\ }\textbf {\bibinfo {volume} {96}},\ \bibinfo
  {pages} {043017} (\bibinfo {year} {2017})}\BibitemShut {NoStop}%
\bibitem [{\citenamefont {Aprile}\ \emph {et~al.}(2019)\citenamefont {Aprile}
  \emph {et~al.}}]{xenon1t_chi_e}%
  \BibitemOpen
  \bibfield  {author} {\bibinfo {author} {\bibfnamefont {E.}~\bibnamefont
  {Aprile}} \emph {et~al.} (\bibinfo {collaboration} {XENON Collaboration}),\
  }\href {\doibase 10.1103/PhysRevLett.123.251801} {\bibfield  {journal}
  {\bibinfo  {journal} {Phys. Rev. Lett.}\ }\textbf {\bibinfo {volume} {123}},\
  \bibinfo {pages} {251801} (\bibinfo {year} {2019})}\BibitemShut {NoStop}%
\bibitem [{\citenamefont {Cheng}\ \emph {et~al.}(2021)\citenamefont {Cheng}
  \emph {et~al.}}]{pandaxii_chi_e}%
  \BibitemOpen
  \bibfield  {author} {\bibinfo {author} {\bibfnamefont {C.}~\bibnamefont
  {Cheng}} \emph {et~al.} (\bibinfo {collaboration} {PandaX-II
  Collaboration}),\ }\href {\doibase 10.1103/PhysRevLett.126.211803} {\bibfield
   {journal} {\bibinfo  {journal} {Phys. Rev. Lett.}\ }\textbf {\bibinfo
  {volume} {126}},\ \bibinfo {pages} {211803} (\bibinfo {year}
  {2021})}\BibitemShut {NoStop}%
\bibitem [{\citenamefont {Agnes}\ \emph
  {et~al.}(2018{\natexlab{b}})\citenamefont {Agnes} \emph
  {et~al.}}]{darkside_chi_e}%
  \BibitemOpen
  \bibfield  {author} {\bibinfo {author} {\bibfnamefont {P.}~\bibnamefont
  {Agnes}} \emph {et~al.} (\bibinfo {collaboration} {DarkSide Collaboration}),\
  }\href {\doibase 10.1103/PhysRevLett.121.111303} {\bibfield  {journal}
  {\bibinfo  {journal} {Phys. Rev. Lett.}\ }\textbf {\bibinfo {volume} {121}},\
  \bibinfo {pages} {111303} (\bibinfo {year} {2018}{\natexlab{b}})}\BibitemShut
  {NoStop}%
\bibitem [{\citenamefont {Soma}\ \emph {et~al.}(2016)\citenamefont {Soma} \emph
  {et~al.}}]{soma2016}%
  \BibitemOpen
  \bibfield  {author} {\bibinfo {author} {\bibfnamefont {A.~K.}\ \bibnamefont
  {Soma}} \emph {et~al.},\ }\href {\doibase 10.1016/j.nima.2016.08.044}
  {\bibfield  {journal} {\bibinfo  {journal} {Nucl. Instrum. Methods Phys.
  Res., Sect. A}\ }\textbf {\bibinfo {volume} {836}},\ \bibinfo {pages} {67 }
  (\bibinfo {year} {2016})}\BibitemShut {NoStop}%
\bibitem [{\citenamefont {Cheng}\ \emph {et~al.}(2017)\citenamefont {Cheng}
  \emph {et~al.}}]{cjpl}%
  \BibitemOpen
  \bibfield  {author} {\bibinfo {author} {\bibfnamefont {J.~P.}\ \bibnamefont
  {Cheng}} \emph {et~al.},\ }\href {\doibase
  10.1146/annurev-nucl-102115-044842} {\bibfield  {journal} {\bibinfo
  {journal} {Annu. Rev. Nucl. Part. Sci.}\ }\textbf {\bibinfo {volume} {67}},\
  \bibinfo {pages} {231} (\bibinfo {year} {2017})}\BibitemShut {NoStop}%
\bibitem [{\citenamefont {She}\ \emph {et~al.}(2020)\citenamefont {She} \emph
  {et~al.}}]{cdexdarkphoton}%
  \BibitemOpen
  \bibfield  {author} {\bibinfo {author} {\bibfnamefont {Z.}~\bibnamefont
  {She}} \emph {et~al.} (\bibinfo {collaboration} {CDEX Collaboration}),\
  }\href {\doibase 10.1103/PhysRevLett.124.111301} {\bibfield  {journal}
  {\bibinfo  {journal} {Phys. Rev. Lett.}\ }\textbf {\bibinfo {volume} {124}},\
  \bibinfo {pages} {111301} (\bibinfo {year} {2020})}\BibitemShut {NoStop}%
\bibitem [{\citenamefont {Xu}\ \emph {et~al.}(2022)\citenamefont {Xu} \emph
  {et~al.}}]{cdexcrdm}%
  \BibitemOpen
  \bibfield  {author} {\bibinfo {author} {\bibfnamefont {R.}~\bibnamefont {Xu}}
  \emph {et~al.} (\bibinfo {collaboration} {CDEX Collaboration}),\ }\href
  {\doibase 10.1103/PhysRevD.106.052008} {\bibfield  {journal} {\bibinfo
  {journal} {Phys. Rev. D}\ }\textbf {\bibinfo {volume} {106}},\ \bibinfo
  {pages} {052008} (\bibinfo {year} {2022})}\BibitemShut {NoStop}%
\bibitem [{\citenamefont {Essig}\ \emph {et~al.}(2012)\citenamefont {Essig},
  \citenamefont {Mardon},\ and\ \citenamefont {Volansky}}]{Ionization}%
  \BibitemOpen
  \bibfield  {author} {\bibinfo {author} {\bibfnamefont {R.}~\bibnamefont
  {Essig}}, \bibinfo {author} {\bibfnamefont {J.}~\bibnamefont {Mardon}}, \
  and\ \bibinfo {author} {\bibfnamefont {T.}~\bibnamefont {Volansky}},\ }\href
  {\doibase 10.1103/PhysRevD.85.076007} {\bibfield  {journal} {\bibinfo
  {journal} {Phys. Rev. D}\ }\textbf {\bibinfo {volume} {85}},\ \bibinfo
  {pages} {076007} (\bibinfo {year} {2012})}\BibitemShut {NoStop}%
\bibitem [{\citenamefont {Graham}\ \emph {et~al.}(2012)\citenamefont {Graham}
  \emph {et~al.}}]{BandGap}%
  \BibitemOpen
  \bibfield  {author} {\bibinfo {author} {\bibfnamefont {P.~W.}\ \bibnamefont
  {Graham}} \emph {et~al.},\ }\href {\doibase
  https://doi.org/10.1016/j.dark.2012.09.001} {\bibfield  {journal} {\bibinfo
  {journal} {Phys. Dark Universe}\ }\textbf {\bibinfo {volume} {1}},\ \bibinfo
  {pages} {32} (\bibinfo {year} {2012})}\BibitemShut {NoStop}%
\bibitem [{\citenamefont {Bunge}\ \emph {et~al.}(1993)\citenamefont {Bunge},
  \citenamefont {Barrientos},\ and\ \citenamefont {Bunge}}]{AtomWaveFunctions}%
  \BibitemOpen
  \bibfield  {author} {\bibinfo {author} {\bibfnamefont {C.}~\bibnamefont
  {Bunge}}, \bibinfo {author} {\bibfnamefont {J.}~\bibnamefont {Barrientos}}, \
  and\ \bibinfo {author} {\bibfnamefont {A.}~\bibnamefont {Bunge}},\ }\href
  {\doibase https://doi.org/10.1006/adnd.1993.1003} {\bibfield  {journal}
  {\bibinfo  {journal} {At. Data Nucl. Data Tables}\ }\textbf {\bibinfo
  {volume} {53}},\ \bibinfo {pages} {113} (\bibinfo {year} {1993})}\BibitemShut
  {NoStop}%
\bibitem [{\citenamefont {Griffin}\ \emph {et~al.}(2021)\citenamefont
  {Griffin}, \citenamefont {Inzani}, \citenamefont {Trickle}, \citenamefont
  {Zhang},\ and\ \citenamefont {Zurek}}]{ExceedDM_PRD}%
  \BibitemOpen
  \bibfield  {author} {\bibinfo {author} {\bibfnamefont {S.~M.}\ \bibnamefont
  {Griffin}}, \bibinfo {author} {\bibfnamefont {K.}~\bibnamefont {Inzani}},
  \bibinfo {author} {\bibfnamefont {T.}~\bibnamefont {Trickle}}, \bibinfo
  {author} {\bibfnamefont {Z.}~\bibnamefont {Zhang}}, \ and\ \bibinfo {author}
  {\bibfnamefont {K.~M.}\ \bibnamefont {Zurek}},\ }\href {\doibase
  10.1103/PhysRevD.104.095015} {\bibfield  {journal} {\bibinfo  {journal}
  {Phys. Rev. D}\ }\textbf {\bibinfo {volume} {104}},\ \bibinfo {pages}
  {095015} (\bibinfo {year} {2021})}\BibitemShut {NoStop}%
\bibitem [{\citenamefont {Essig}\ \emph {et~al.}(2016)\citenamefont {Essig},
  \citenamefont {Fern{\'a}ndez-Serra}, \citenamefont {Mardon}, \citenamefont
  {Soto}, \citenamefont {Volansky},\ and\ \citenamefont
  {Yu}}]{chi_e_scattering_rate}%
  \BibitemOpen
  \bibfield  {author} {\bibinfo {author} {\bibfnamefont {R.}~\bibnamefont
  {Essig}}, \bibinfo {author} {\bibfnamefont {M.}~\bibnamefont
  {Fern{\'a}ndez-Serra}}, \bibinfo {author} {\bibfnamefont {J.}~\bibnamefont
  {Mardon}}, \bibinfo {author} {\bibfnamefont {A.}~\bibnamefont {Soto}},
  \bibinfo {author} {\bibfnamefont {T.}~\bibnamefont {Volansky}}, \ and\
  \bibinfo {author} {\bibfnamefont {T.-T.}\ \bibnamefont {Yu}},\ }\href
  {\doibase 10.1007/JHEP05(2016)046} {\bibfield  {journal} {\bibinfo  {journal}
  {J. High Energy Phys.}\ }\textbf {\bibinfo {volume} {05}},\ \bibinfo {pages}
  {46} (\bibinfo {year} {2016})}\BibitemShut {NoStop}%
\bibitem [{\citenamefont {Li}\ \emph {et~al.}(2015)\citenamefont {Li},
  \citenamefont {Miao},\ and\ \citenamefont {Zhou}}]{light_mediator}%
  \BibitemOpen
  \bibfield  {author} {\bibinfo {author} {\bibfnamefont {T.}~\bibnamefont
  {Li}}, \bibinfo {author} {\bibfnamefont {S.}~\bibnamefont {Miao}}, \ and\
  \bibinfo {author} {\bibfnamefont {Y.-F.}\ \bibnamefont {Zhou}},\ }\href
  {\doibase 10.1088/1475-7516/2015/03/032} {\bibfield  {journal} {\bibinfo
  {journal} {J. Cosmol. Astropart. Phys.}\ }\textbf {\bibinfo {volume} {03}},\
  \bibinfo {pages} {032} (\bibinfo {year} {2015})}\BibitemShut {NoStop}%
\bibitem [{\citenamefont {Lewin}\ and\ \citenamefont {Smith}(1996)}]{SHM}%
  \BibitemOpen
  \bibfield  {author} {\bibinfo {author} {\bibfnamefont {J.~D.}\ \bibnamefont
  {Lewin}}\ and\ \bibinfo {author} {\bibfnamefont {P.~F.}\ \bibnamefont
  {Smith}},\ }\href {\doibase 10.1016/S0927-6505(96)00047-3} {\bibfield
  {journal} {\bibinfo  {journal} {Astropart. Phys.}\ }\textbf {\bibinfo
  {volume} {6}},\ \bibinfo {pages} {87 } (\bibinfo {year} {1996})}\BibitemShut
  {NoStop}%
\bibitem [{\citenamefont {Smith}\ \emph {et~al.}(2007)\citenamefont {Smith}
  \emph {et~al.}}]{escapevelocity}%
  \BibitemOpen
  \bibfield  {author} {\bibinfo {author} {\bibfnamefont {M.~C.}\ \bibnamefont
  {Smith}} \emph {et~al.},\ }\href {\doibase 10.1111/j.1365-2966.2007.11964.x}
  {\bibfield  {journal} {\bibinfo  {journal} {Mon. Not. R. Astron. Soc.}\
  }\textbf {\bibinfo {volume} {379}},\ \bibinfo {pages} {755} (\bibinfo {year}
  {2007})}\BibitemShut {NoStop}%
\bibitem [{\citenamefont {Lee}\ \emph {et~al.}(2015)\citenamefont {Lee},
  \citenamefont {Lisanti}, \citenamefont {Mishra-Sharma},\ and\ \citenamefont
  {Safdi}}]{SemiAnalytic1}%
  \BibitemOpen
  \bibfield  {author} {\bibinfo {author} {\bibfnamefont {S.~K.}\ \bibnamefont
  {Lee}}, \bibinfo {author} {\bibfnamefont {M.}~\bibnamefont {Lisanti}},
  \bibinfo {author} {\bibfnamefont {S.}~\bibnamefont {Mishra-Sharma}}, \ and\
  \bibinfo {author} {\bibfnamefont {B.~R.}\ \bibnamefont {Safdi}},\ }\href
  {\doibase 10.1103/PhysRevD.92.083517} {\bibfield  {journal} {\bibinfo
  {journal} {Phys. Rev. D}\ }\textbf {\bibinfo {volume} {92}},\ \bibinfo
  {pages} {083517} (\bibinfo {year} {2015})}\BibitemShut {NoStop}%
\bibitem [{\citenamefont {Derenzo}\ \emph {et~al.}(2017)\citenamefont
  {Derenzo}, \citenamefont {Essig}, \citenamefont {Massari}, \citenamefont
  {Soto},\ and\ \citenamefont {Yu}}]{QEDark}%
  \BibitemOpen
  \bibfield  {author} {\bibinfo {author} {\bibfnamefont {S.}~\bibnamefont
  {Derenzo}}, \bibinfo {author} {\bibfnamefont {R.}~\bibnamefont {Essig}},
  \bibinfo {author} {\bibfnamefont {A.}~\bibnamefont {Massari}}, \bibinfo
  {author} {\bibfnamefont {A.}~\bibnamefont {Soto}}, \ and\ \bibinfo {author}
  {\bibfnamefont {T.-T.}\ \bibnamefont {Yu}},\ }\href {\doibase
  10.1103/PhysRevD.96.016026} {\bibfield  {journal} {\bibinfo  {journal} {Phys.
  Rev. D}\ }\textbf {\bibinfo {volume} {96}},\ \bibinfo {pages} {016026}
  (\bibinfo {year} {2017})}\BibitemShut {NoStop}%
\bibitem [{\citenamefont {Hochberg}\ \emph {et~al.}(2021)\citenamefont
  {Hochberg}, \citenamefont {Kahn}, \citenamefont {Kurinsky}, \citenamefont
  {Lehmann}, \citenamefont {Yu},\ and\ \citenamefont
  {Berggren}}]{dielectric_scattering}%
  \BibitemOpen
  \bibfield  {author} {\bibinfo {author} {\bibfnamefont {Y.}~\bibnamefont
  {Hochberg}}, \bibinfo {author} {\bibfnamefont {Y.}~\bibnamefont {Kahn}},
  \bibinfo {author} {\bibfnamefont {N.}~\bibnamefont {Kurinsky}}, \bibinfo
  {author} {\bibfnamefont {B.~V.}\ \bibnamefont {Lehmann}}, \bibinfo {author}
  {\bibfnamefont {T.~C.}\ \bibnamefont {Yu}}, \ and\ \bibinfo {author}
  {\bibfnamefont {K.~K.}\ \bibnamefont {Berggren}},\ }\href {\doibase
  10.1103/PhysRevLett.127.151802} {\bibfield  {journal} {\bibinfo  {journal}
  {Phys. Rev. Lett.}\ }\textbf {\bibinfo {volume} {127}},\ \bibinfo {pages}
  {151802} (\bibinfo {year} {2021})}\BibitemShut {NoStop}%
\bibitem [{\citenamefont {Knapen}\ \emph {et~al.}(2021)\citenamefont {Knapen},
  \citenamefont {Kozaczuk},\ and\ \citenamefont {Lin}}]{elf}%
  \BibitemOpen
  \bibfield  {author} {\bibinfo {author} {\bibfnamefont {S.}~\bibnamefont
  {Knapen}}, \bibinfo {author} {\bibfnamefont {J.}~\bibnamefont {Kozaczuk}}, \
  and\ \bibinfo {author} {\bibfnamefont {T.}~\bibnamefont {Lin}},\ }\href
  {\doibase 10.1103/PhysRevD.104.015031} {\bibfield  {journal} {\bibinfo
  {journal} {Phys. Rev. D}\ }\textbf {\bibinfo {volume} {104}},\ \bibinfo
  {pages} {015031} (\bibinfo {year} {2021})}\BibitemShut {NoStop}%
\bibitem [{\citenamefont {Knapen}\ \emph {et~al.}(2022)\citenamefont {Knapen},
  \citenamefont {Kozaczuk},\ and\ \citenamefont {Lin}}]{darkelf}%
  \BibitemOpen
  \bibfield  {author} {\bibinfo {author} {\bibfnamefont {S.}~\bibnamefont
  {Knapen}}, \bibinfo {author} {\bibfnamefont {J.}~\bibnamefont {Kozaczuk}}, \
  and\ \bibinfo {author} {\bibfnamefont {T.}~\bibnamefont {Lin}},\ }\href
  {\doibase 10.1103/PhysRevD.105.015014} {\bibfield  {journal} {\bibinfo
  {journal} {Phys. Rev. D}\ }\textbf {\bibinfo {volume} {105}},\ \bibinfo
  {pages} {015014} (\bibinfo {year} {2022})}\BibitemShut {NoStop}%
\bibitem [{\citenamefont {Gelmini}\ \emph {et~al.}(2020)\citenamefont
  {Gelmini}, \citenamefont {Takhistov},\ and\ \citenamefont
  {Vitagliano}}]{scalarinteraction}%
  \BibitemOpen
  \bibfield  {author} {\bibinfo {author} {\bibfnamefont {G.~B.}\ \bibnamefont
  {Gelmini}}, \bibinfo {author} {\bibfnamefont {V.}~\bibnamefont {Takhistov}},
  \ and\ \bibinfo {author} {\bibfnamefont {E.}~\bibnamefont {Vitagliano}},\
  }\href {\doibase https://doi.org/10.1016/j.physletb.2020.135779} {\bibfield
  {journal} {\bibinfo  {journal} {Phys. Lett. B}\ }\textbf {\bibinfo {volume}
  {809}},\ \bibinfo {pages} {135779} (\bibinfo {year} {2020})}\BibitemShut
  {NoStop}%
\bibitem [{\citenamefont {Liang}\ \emph {et~al.}(2019)\citenamefont {Liang},
  \citenamefont {Zhang}, \citenamefont {Zhang},\ and\ \citenamefont
  {Zheng}}]{AE}%
  \BibitemOpen
  \bibfield  {author} {\bibinfo {author} {\bibfnamefont {Z.~L.}\ \bibnamefont
  {Liang}}, \bibinfo {author} {\bibfnamefont {L.}~\bibnamefont {Zhang}},
  \bibinfo {author} {\bibfnamefont {P.}~\bibnamefont {Zhang}}, \ and\ \bibinfo
  {author} {\bibfnamefont {F.~W.}\ \bibnamefont {Zheng}},\ }\href {\doibase
  10.1007/JHEP01(2019)149} {\bibfield  {journal} {\bibinfo  {journal} {J. High
  Energy Phys.}\ }\textbf {\bibinfo {volume} {01}},\ \bibinfo {pages} {149}
  (\bibinfo {year} {2019})}\BibitemShut {NoStop}%
\bibitem [{\citenamefont {Trickle}\ and\ \citenamefont
  {kinzani}(2022)}]{ExceedDM020}%
  \BibitemOpen
  \bibfield  {author} {\bibinfo {author} {\bibfnamefont {T.}~\bibnamefont
  {Trickle}}\ and\ \bibinfo {author} {\bibnamefont {kinzani}},\ }\href
  {\doibase 10.5281/zenodo.6097642} {\enquote {\bibinfo {title}
  {{tanner-trickle/EXCEED-DM: EXCEED-DM-v0.3.0}},}\ } (\bibinfo {year}
  {2022})\BibitemShut {NoStop}%
\bibitem [{\citenamefont {Pandey}\ \emph {et~al.}(2020)\citenamefont {Pandey},
  \citenamefont {Singh}, \citenamefont {Wu}, \citenamefont {Chen},
  \citenamefont {Chi}, \citenamefont {Hsieh}, \citenamefont {Liu},\ and\
  \citenamefont {Wong}}]{effectivefieldtheory}%
  \BibitemOpen
  \bibfield  {author} {\bibinfo {author} {\bibfnamefont {M.~K.}\ \bibnamefont
  {Pandey}}, \bibinfo {author} {\bibfnamefont {L.}~\bibnamefont {Singh}},
  \bibinfo {author} {\bibfnamefont {C.-P.}\ \bibnamefont {Wu}}, \bibinfo
  {author} {\bibfnamefont {J.-W.}\ \bibnamefont {Chen}}, \bibinfo {author}
  {\bibfnamefont {H.-C.}\ \bibnamefont {Chi}}, \bibinfo {author} {\bibfnamefont
  {C.-C.}\ \bibnamefont {Hsieh}}, \bibinfo {author} {\bibfnamefont {C.-P.}\
  \bibnamefont {Liu}}, \ and\ \bibinfo {author} {\bibfnamefont {H.~T.}\
  \bibnamefont {Wong}},\ }\href {\doibase 10.1103/PhysRevD.102.123025}
  {\bibfield  {journal} {\bibinfo  {journal} {Phys. Rev. D}\ }\textbf {\bibinfo
  {volume} {102}},\ \bibinfo {pages} {123025} (\bibinfo {year}
  {2020})}\BibitemShut {NoStop}%
\bibitem [{\citenamefont {Li}\ \emph {et~al.}(2014)\citenamefont {Li} \emph
  {et~al.}}]{Li:2014a}%
  \BibitemOpen
  \bibfield  {author} {\bibinfo {author} {\bibfnamefont {H.~B.}\ \bibnamefont
  {Li}} \emph {et~al.},\ }\href {\doibase
  https://doi.org/10.1016/j.astropartphys.2014.02.005} {\bibfield  {journal}
  {\bibinfo  {journal} {Astropart. Phys.}\ }\textbf {\bibinfo {volume} {56}},\
  \bibinfo {pages} {1} (\bibinfo {year} {2014})}\BibitemShut {NoStop}%
\bibitem [{\citenamefont {Yang}\ \emph
  {et~al.}(2018{\natexlab{b}})\citenamefont {Yang} \emph
  {et~al.}}]{Yang:2018a}%
  \BibitemOpen
  \bibfield  {author} {\bibinfo {author} {\bibfnamefont {L.~T.}\ \bibnamefont
  {Yang}} \emph {et~al.},\ }\href {\doibase
  https://doi.org/10.1016/j.nima.2017.12.078} {\bibfield  {journal} {\bibinfo
  {journal} {Nucl. Instrum. Methods Phys. Res., Sect. A}\ }\textbf {\bibinfo
  {volume} {886}},\ \bibinfo {pages} {13} (\bibinfo {year}
  {2018}{\natexlab{b}})}\BibitemShut {NoStop}%
\bibitem [{\citenamefont {Feldman}\ and\ \citenamefont
  {Cousins}(1998)}]{chisquare}%
  \BibitemOpen
  \bibfield  {author} {\bibinfo {author} {\bibfnamefont {G.~J.}\ \bibnamefont
  {Feldman}}\ and\ \bibinfo {author} {\bibfnamefont {R.~D.}\ \bibnamefont
  {Cousins}},\ }\href {\doibase 10.1103/PhysRevD.57.3873} {\bibfield  {journal}
  {\bibinfo  {journal} {Phys. Rev. D}\ }\textbf {\bibinfo {volume} {57}},\
  \bibinfo {pages} {3873} (\bibinfo {year} {1998})}\BibitemShut {NoStop}%
\bibitem [{\citenamefont {Emken}\ \emph {et~al.}(2019)\citenamefont {Emken},
  \citenamefont {Essig}, \citenamefont {Kouvaris},\ and\ \citenamefont
  {Sholapurkar}}]{earthshielding_chi_e}%
  \BibitemOpen
  \bibfield  {author} {\bibinfo {author} {\bibfnamefont {T.}~\bibnamefont
  {Emken}}, \bibinfo {author} {\bibfnamefont {R.}~\bibnamefont {Essig}},
  \bibinfo {author} {\bibfnamefont {C.}~\bibnamefont {Kouvaris}}, \ and\
  \bibinfo {author} {\bibfnamefont {M.}~\bibnamefont {Sholapurkar}},\ }\href
  {\doibase 10.1088/1475-7516/2019/09/070} {\bibfield  {journal} {\bibinfo
  {journal} {J. Cosmol. Astropart. Phys.}\ }\textbf {\bibinfo {volume} {09}},\
  \bibinfo {pages} {070} (\bibinfo {year} {2019})}\BibitemShut {NoStop}%
\bibitem [{\citenamefont {Abramoff}\ \emph {et~al.}(2019)\citenamefont
  {Abramoff} \emph {et~al.}}]{protosensei_chi_e}%
  \BibitemOpen
  \bibfield  {author} {\bibinfo {author} {\bibfnamefont {O.}~\bibnamefont
  {Abramoff}} \emph {et~al.} (\bibinfo {collaboration} {SENSEI
  Collaboration}),\ }\href {\doibase 10.1103/PhysRevLett.122.161801} {\bibfield
   {journal} {\bibinfo  {journal} {Phys. Rev. Lett.}\ }\textbf {\bibinfo
  {volume} {122}},\ \bibinfo {pages} {161801} (\bibinfo {year}
  {2019})}\BibitemShut {NoStop}%
\bibitem [{\citenamefont {Agnese}\ \emph
  {et~al.}(2018{\natexlab{b}})\citenamefont {Agnese} \emph
  {et~al.}}]{cdmshvev_chi_e_r1}%
  \BibitemOpen
  \bibfield  {author} {\bibinfo {author} {\bibfnamefont {R.}~\bibnamefont
  {Agnese}} \emph {et~al.} (\bibinfo {collaboration} {SuperCDMS
  Collaboration}),\ }\href {\doibase 10.1103/PhysRevLett.121.051301} {\bibfield
   {journal} {\bibinfo  {journal} {Phys. Rev. Lett.}\ }\textbf {\bibinfo
  {volume} {121}},\ \bibinfo {pages} {051301} (\bibinfo {year}
  {2018}{\natexlab{b}})}\BibitemShut {NoStop}%
\end{thebibliography}%

\end{document}